\documentclass[12pt]{amsart}
\usepackage[
  top=1.15in,
  bottom=1.15in,
  left=1.25in,
  right=1.25in,
  heightrounded,
]{geometry}
\usepackage{txfonts}

\theoremstyle{plain}
\newtheorem{theorem}{Theorem}
\newtheorem{lemma}[theorem]{Lemma}
\newtheorem{proposition}[theorem]{Proposition}

\newtheorem{definition}[theorem]{Definition}

\begin{document}

\title[Fractal Sets and the Kadomtsev--Petviashvili Equation]{Distributions Supported on Fractal Sets and Solutions to the Kadomtsev--Petviashvili Equation}
\author{Patrik V. Nabelek}
\date{\today}

\begin{abstract}

In this note we will discuss a potentially interesting extension of some recent results on primitive solutions to completely integrable partial differential equations.
We will discuss a family distributions that are holomorphic on the Riemann sphere except on the singular sets homeomorphic to a Cantor set or Sierpinski gasket.
These distributions allow us to produce solutions to the Kadomtsev--Petviashvili equation.
These distributions are limits of families of rational functions that can also be associated with holomorphic line bundles on surfaces with a finite number of doubly degenerate singular points.
We conjecture that a subset of these distributions can be used to formulate a definition of a holomorphic line bundle on some surfaces that are homeomorphic to spheres except where they become doubly degenerate on singular sets homeomorphic to a Cantor set or Sierpinski gasket.

\end{abstract}

\maketitle
\section{Introduction}
In this note we will discuss examples of distributions in the sense of Schwartz that we believe are promising for the study of the solutions to the Kadomtsev--Petviashvili.
These distributions allow the computation of solutions Kadomtsev--Petviashvili (KP) equation and the Korteweg--de Vries (KdV) equation.
The main mathematical method used in this note is the nonlocal dbar problem.
The nonlocal dbar problem was originally introduced by Ablowitz, Bar-Yaacov, and Fokas \cite{ABF83} to formulate the inverse scattering transform for the KP equation.
This method was further generalized by Zakharov and Manakov \cite{ZM85} to other (2+1)D completely integrable systems.
Recently, the nonlocal dbar problem has been used to study new classes of solutions to the KdV equation called primitive solutions \cite{DNZZ20,DZZ16,GGJM18,NZZ19,N20a,ZDZ16,ZZD16} and connections to the KP equation were discussed in \cite{N20b}.

The distributions introduced in this note are singular on a Cantor set or Sierpinski gasket, where the singularities can potentially correspond to nonlocal identifications along two copies of a Cantor set or Sierpinski gasket on the Riemann sphere.
These spaces of distributions are formed using the nonlocal dbar problem, and a limit of rational functions on the plane as the number of poles diverges to infinity.

\subsection{Prerequisite Definitions}

All distributions that appear in this note will be defined with respect to compactly supported smooth test functions.

\begin{definition} \label{defQnQ}
Suppose $\mathcal{Q}_n \subset \mathbb{C}$ is a sequence of complex point sets such that $\mathcal{Q}_n \subset \mathcal{Q}_{n+1}$ and
$$\mathcal{Q} = \overline{\lim_{n \to \infty}} \mathcal{Q}_n := \overline{\{x \in \mathcal{Q}_n : n = 0,1,2,\dots\}}$$
such that
$$\lim_{n \to \infty} \frac{1}{|\mathcal{Q}_n|} \sum_{\lambda_n \in \mathcal{Q}_n} \delta(\lambda - \lambda_n) d\lambda = d \mu_{\mathcal{Q}} (\lambda)$$
where $\mu_{\mathcal{Q}} (\lambda)$ is the probability measure on $\mathbb{C}$ that is supported on $\mathcal{Q}$ where it restricts to the uniform probability measure on $\mathcal{Q}$.  
\end{definition}

\begin{definition}
Let $H_\mu(\mathcal{Q})$  be the set of H\"{o}lder continuous functions on $\mathcal{Q}$ with H\"{o}lder coefficient $0<\mu<1$.
\end{definition}

\begin{definition} The following definitions are either classical constructions of Cantor \cite{C1883} and Sierpinski \cite{S1915} or explicit:
\begin{enumerate}
\item The $n$-th step of the Cantor iteration for the Cantor middle $\epsilon$ set $\mathcal{C}_n \subset \mathbb{C}$.
\item The point set $\mathcal{P}_n \subset \mathcal{C}_n$ of the endpoints of the intervals making up $\mathcal{C}_n$.
\item The $n$-th step $\mathcal{S}_n \subset \mathbb{C}$ of the iteration used to construct the Sierpinski gasket. 
\item The vertex sets $\mathcal{V}_n \subset \mathcal{S}_n$ of vertices of the triangles making up $\mathcal{S}_n$.
\item The Cantor set can be defined by
$$\mathcal{C} = \bigcap_{n=0}^\infty \mathcal{C}_n.$$
\item The Sierpinski gasket can be defined by
$$\mathcal{S} = \bigcap_{n=0}^\infty \mathcal{S}_n.$$
\end{enumerate}
\end{definition}

\begin{proposition}
The following are true:
\begin{enumerate}
\item The point sets $\mathcal{P}_n$ and $\mathcal{C}$ are related by
$$\overline{\lim_{n \to \infty}} \mathcal{P}_n = \mathcal{C}$$
and satisfy the necessary properties from definition \ref{defQnQ} to be a choice of $\mathcal{Q}_n$ and $\mathcal{Q}$ respectively.
\item The point sets $\mathcal{V}_n$ and $\mathcal{S}$ are related by
$$\mathcal{S} = \overline{\lim_{n \to \infty}} \mathcal{V}_n$$
and satisfy the necessary properties from definition \ref{defQnQ} to be a choice of $\mathcal{Q}_n$ and $\mathcal{Q}$ respectively.
\end{enumerate}
\end{proposition}

\section{The Nonlocal Dbar Problem and Spaces of Distributions}

\begin{definition}
We will use $\mathcal{Q}_n$ to refer to either $\mathcal{P}_n$ or $\mathcal{V}_n$ and we will let $$Q = \overline{\lim_{n \to \infty}} \mathcal{Q}_n.$$
We will consider some explicit spaces of compactly supported distributions defined with respect to smooth compactly supported test functions as follows:
\begin{enumerate}
\item From $\mathcal{Q}_n$ and a list of numbers $a_j$ we determine the following distributions
\begin{align*} & \chi_n(\lambda) = 1 + \frac{1}{\pi |\mathcal{Q}_n|} \sum_{\lambda_j \in \mathcal{Q}_n} \frac{a_j}{\lambda - \lambda_j} \\
& \chi(\lambda) = \lim_{n \to \infty} \chi_n(\lambda).\end{align*}
(at this point we make no comment on the existence of the limiting distribution)
\item Let $\phi$ be some isometry of $\mathbb{C}$ such that $\phi: \mathcal{Q} \to \mathcal{Q}' \subset \mathbb{C}$ where $\mathcal{Q}$ and $\mathcal{Q}'$ are disjoint.
If we enforce the nonlocal conditions
$$a_j = r(\lambda_j)\chi_n(\phi(\lambda_j))$$
for all $j$ where $r \in H_\mu(\mathcal{Q})$ then the rational functions $\chi_n$ are uniquely defined by the above conditions.
\end{enumerate}
\end{definition}

The distributions $\chi_n(\lambda)$ can be identified with rational functions by definition, but the distributions $\chi(\lambda)$ cannot be represented by rational functions. An equivalent construction can also be made using functions and measures instead of distributions and rational functions.
In the previous definition, the closure of the rational functions in the topology of uniform convergence in compact sets that do not intersect the singular sets of the surface can intuitively be formulated formally using limits of rational functions, and can be made rigorous using Schwartz's theory of distributions.

\begin{proposition}
Let $\mathcal{Q}_n \supset \mathcal{Q}_{n-1}$ and $\mathcal{R}_n \supset \mathcal{R}_{n-1}$ be two sequences of point sets such that
$$ \overline{\lim_{n \to \infty} \mathcal{Q}_n}=\mathcal{Q}, \quad \overline{\lim_{n \to \infty} \mathcal{R}_n} = \mathcal{Q}'$$
but so that the point measures on $\mathcal{Q}_n$ and $\mathcal{R}_n$ limit to measures on $\mathcal{Q}$ and $\mathcal{Q}'$, and $\phi(\mathcal{Q}_n)$ and $\mathcal{R}_n$ are disjoint. Now consider the distributions of the form
\begin{align*} &  \tilde \chi_n(\lambda) = 1 + \frac{1}{\pi |\mathcal{Q}_n|}\left(\sum_{\lambda_j \in \mathcal{Q}_n} \frac{a_j}{\lambda - \lambda_j}\right) + \frac{1}{\pi |\mathcal{R}_n|}\left(\sum_{\mu_k \in \mathcal{R}_n} \frac{b_j}{\lambda - \mu_k} \right) \\
& \tilde \chi(\lambda) = \lim_{n \to \infty} \tilde \chi_n(\lambda)
\end{align*}
where $a_j$ and $b_j$ are bounded sequences associated to the common points of $\mathcal{Q}_n$ (at this point we make no comment on the existence of the limiting distribution).
If we assume the following nonlocal conditions on the poles
$$a_j = r_1(\lambda_j) \chi_n(\phi(\lambda_j)), \quad b_k =r_2(\phi^{-1}(\mu_k)) \chi_n(\phi^{-1}(\mu_k))$$
for all $j$ where $r_1,r_2 \in H_{\mu}(\mathcal{Q})$ then the rational functions $\tilde \chi_n(\lambda)$ are uniquely defined by the above conditions. 
\end{proposition}

The following lemma will be important, because it allows us to define the limit of the rational functions $\chi_n(\lambda)$ and $\tilde \chi_n(\lambda)$ as $n \to \infty$. This lemma allows us to study functions on $\mathcal{Q}$ that solve singular integral equations instead of the nonlocal dbar problem. 

\begin{lemma} \label{lma} The rational functions $\chi_n$ and $\tilde \chi_n$ solve the nonlocal dbar problems
\begin{equation} \label{eqdbar} \frac{\partial \breve{\chi}}{\partial \lambda}(\lambda) = R(\lambda) \breve{\chi}(\phi(\lambda)) \end{equation}
where
\begin{equation} R(\lambda) = \frac{1}{|\mathcal{Q}_n|} \sum_{\lambda_j \in \mathcal{Q}_n} r(\lambda_j) \delta(\lambda - \lambda_j), \end{equation}
or
\begin{equation} R(\lambda) = \frac{1}{|\mathcal{Q}_n|} \sum_{\lambda_j \in \mathcal{Q}_n} r_1(\lambda_j) \delta(\lambda - \lambda_j) + \frac{1}{|\mathcal{R}_n|}\sum_{\mu_k \in \mathcal{R}_n} r_2(\phi^{-1}(\mu_k)) \delta(\lambda - \mu_k) \end{equation}
and $\breve \chi(\lambda) \to 1$ as $\lambda \to \infty$.
Moreover, if we suppose that $r,r_1, r_2\in H_\mu(\mathcal{Q})$  then $R(\lambda)$ limits to
\begin{equation} \label{eqR1} R(\lambda) = \int_{\mathcal{Q}} r(s) \delta(\lambda-s) d\mu_\mathcal{Q}(s),\end{equation}
or
\begin{equation} \label{eqR2} R(\lambda) = \int_{\mathcal{Q}} r_1(s) \delta(\lambda-s) + r_2(s) \delta(\lambda-\phi(s)) d\mu_{\mathcal{Q}}(s) \end{equation}
respectively as $n \to \infty$.
\end{lemma}

Because of the assumption that the uniform probability measure $d\mu_{\mathcal{Q}}$ exists as the limit of delta measure, the proof of this lemma is a simply application of the theory of distributions. One just needs to apply the distributions to test functions, and then take the limit.
The proof is left to the reader.

The functions $\chi_n(\lambda)$ and $\tilde \chi_n(\lambda)$ determine a holomorphic line bundle on a singular rational curve \cite{NZZ19}. As $n \to \infty$, the poles coalesce into the singular sets $\mathcal{Q}$ and $\mathcal{Q}'$ that are identified via the nonlocal identification. For the case of finite and infinite gap solutions to the the KdV equation, we can take $\mathcal{Q}$ and $\mathcal{Q}'$ to be intervals, and the limit gives the finite gap solutions to the KdV equation as primitive solutions to the KdV equation.

The nonlocal dbar can instead be formulated as a system of two (local) singular integral equations. The justification is a simple implication of the constructions in \cite{DZZ16,DNZZ20,GGJM18,NZZ19,N20a,N20b}. The limiting procedure used in \cite{DZZ16} that was the inspiration for further results in \cite{DNZZ20,NZZ19,N20a,N20b} is deterministic.
However, this limit is ineffective for some numerical and statistical calculations.

In \cite{GGJM18} a random discrete soliton amplitude spectrum that leads to a Riemann--Stiljes integration is used and gives an alternative way of rigorously defining the functions $r(s)$, $r_1(s)$ and $r_2(s)$ in the singular integral equations in the case of a primitive solution to the KdV equation.
The universality of the Riemann--Stiljes integral for any choice of partition generating it, also guarantees the definiteness of the limiting distribution. 

In the case considered in this note, these methods lead to the following theorem.
\begin{theorem}\label{thm}
The limiting distributions from Lemma \ref{lma} can be written in the forms
\begin{align} &\label{eqchi1} \chi(\lambda) = 1 + \frac{1}{\pi} \int_{\mathcal{Q}} \frac{f(s)}{\lambda-s} d\mu_{\mathcal{Q}}(s)  \\
& \label{eqtchi1} \tilde \chi(\lambda) = 1 + \frac{1}{\pi} \int_{\mathcal{Q}} \left(\frac{f_1(s)}{\lambda-s}  +  \frac{f_2(s)}{\lambda - \phi(s)} \right) d\mu_{\mathcal{Q}}(s). \end{align}
Moreover, the function $f(s)$ solves the integral equation
\begin{equation} \label{ieq1} r(t) = f(t) + \frac{r(t)}{\pi} \int_{\mathcal{Q}} \frac{f(s)}{\phi(t)-s} d \mu_{\mathcal{Q}}(s) \end{equation}
and the functions $f_1(s)$ and $f_2(s)$ solve the system of singular integral equations
\begin{align} \label{ieq2}& r_1(t) = f_1(t) + \frac{r_1(t)}{\pi} \left(\int_{\mathcal{Q}} \frac{f_1(s)}{\phi(t)-s} d \mu_{\mathcal{Q}}(s) + \fint_{\mathcal{Q}} \frac{f_2(s)}{\phi(t) - \phi(s)} d \mu_{\mathcal{Q}}(s)\right)   \\
\label{ieq3}& r_2(t) = f_2(t) + \frac{r_2(t)}{\pi} \left(  \fint_{\mathcal{Q}} \frac{f_1(s)}{t-s} d \mu_{\mathcal{Q}}(s) + \int_{\mathcal{Q}} \frac{f_2(s)}{t-\phi(s)} d \mu_{\mathcal{Q}}(s)  \right)  \end{align}
which determine $f(s)$, $f_1(s)$ and $f_2(s)$ from $r,r_1,r_2\in H_\mu(\mathcal{Q})$. The principle value integrals are defined via the embedding of $\mathcal{C}$ or $\mathcal{S}$ into $\mathbb{C}$.
\end{theorem}

The integral equations \eqref{ieq1}, \eqref{ieq2}, \eqref{ieq3} can be produced immediately using the formal proof of the main theorem from \cite{N20b}. This formal limiting argument leads to well defined integrals because of the assumption on the existance of $d\mu_\mathcal{Q}$ in definition \ref{defQnQ}.

This theorem is proven by plugging \eqref{eqchi1} and \eqref{eqtchi1} into \eqref{eqdbar}, using the functions $R$ of the form \eqref{eqR1} and \eqref{eqR2} respectively.
The nonlocal dbar problem leads to the following integral equation
$$ \breve \chi(\lambda) = 1 + \frac{1}{\pi} \iint_{\mathbb{C}} \frac{R(\zeta)\chi(\phi(\zeta))}{\lambda - \zeta} d^2 \zeta $$
where $d^2 \zeta$ is the usual area form on $\mathbb{C} \equiv \mathbb{R}^2$.
This integral equation comes from combining the inversion formula for the dbar operator and the asymptotic condition $\breve \chi(\lambda) \to 1$ as $\lambda \to \infty$.
The integral equation is determined by its behavior near the singular support, because $\chi(\lambda)$ solves the Cauchy--Riemann equations off of the singular support.
This is because with no singular support, the Cauchy--Riemann equations would imply $\breve \chi(\lambda) = 1$.
Therefore, the singular support is acting as a source for the Cauchy--Riemann equations (in a manner analogous to the way electrostatic charge distributions generate solutions to the Laplace equation).

\begin{definition}
We will use $\mathcal{D}_\bullet'$ to refer to the distribution $\chi(\lambda)$ and $\tilde\chi(\lambda)$ that are determined by the singular integral equations in theorem \ref{thm}.
\end{definition}

It seems likely that distributions $\tilde \chi(\lambda)$ in $\mathcal{D}_\bullet'$ could be used to extend the idea of a holomorphic line bundle on a singular rational curve, due to the interpretations of $\chi_n(\lambda)$ and $\tilde \chi_n(\lambda)$ as giving holomorphic line bundles on singular surfaces \cite{NZZ19}.
The constructions in \cite{NZZ19,N20a} give evidence to the conjecture that the idea of a holomorphic line bundles on surfaces can be extended to the singular surface $\Sigma_{\mathcal{Q}} = S^2/ \left< \sim \right>$ --- which is the topological surface formed by taking $\mathbb{C}$ and identifying $\mathcal{Q}$ with $\mathcal{Q}'$ by $\sim$ via restriction of the isometry $\phi$ to $\mathcal{Q}$ --- using certain choices of $\tilde \chi(\lambda)$, or equivalently to certain choices of $r_1(s) \ge 0$ and $r_2(s) \le 0$.

\begin{definition}
When $\mathcal{Q} = \mathcal{C}$ we call $\Sigma_{\mathcal{Q}}$ a Cantor surface, and when $\mathcal{Q} = \mathcal{S}$ we call $\Sigma_{\mathcal{Q}}$ a Sierpinski surface. We will call $\tilde \chi(\lambda)$ a primitive distribution due to the connection with primitive potentials.
\end{definition}

An explicit link between a notion of holomorphic line bundles on these singular surfaces and the distributions discussed in the previous section would lead to an idea of a Picard group of the Cantor and Sierpinski surfaces.

\begin{definition}
We can form the holomorphic one forms
$$\tilde \omega_n = \tilde \chi_n(\lambda) d \lambda, \quad \tilde \omega = \tilde \chi(\lambda) d \lambda.$$
We can define the space $\mathcal{A}$ of holomorphic differentials on
$$\mathbb{C} \setminus \mathcal{Q}$$ of the form 
$$\omega = \chi(\lambda) d \lambda,$$
and the space $\tilde{\mathcal{A}}$ of holomorphic differentials on
$$\mathbb{C} \setminus (\mathcal{Q} \cup \mathcal{Q}')$$
of the form
$$\tilde \omega = \tilde \chi(\lambda) d \lambda.$$
\end{definition}

It  seems likely that for some choices of functions $r_1(s)$ and $r_2(s)$ these one forms can be interpreted as holomorphic one forms on $\Sigma_{\mathcal{Q}}$.
This interpretation would be important to singularity theory in complex geometry because it would be an example of a complicated singular set for which the surface $\Sigma_{\mathcal{Q}}$ can still be given a large family of holomorphic one forms.
These one forms could also potentially be used to define an idea of a holomorphic line bundle on $\Sigma_{\mathcal{Q}}$.

The conjectures discussed at the end of this section also have evidence in connections between solutions to the nonlocal dbar problem and solutions to the KP equation.

\section{The (2+1)D Kadomtsev--Petviashvili, the (1+1)D Korteweg--de Vries, and the (1+1)D Schr\"{o}dinger Equation.}

Define the phase function
$$\psi(s,x,y,t) = s x + s^2 y +  s^3 t .$$
To produce a solution to the complex KP equation
\begin{equation} \label{eqKP} (4u_t + 6uu_x - u_{xxx})_x -3u_{yy} = 0.\end{equation}
we simply need to assume
$$r_1(s) = e^{\psi(s,x,y,t)-\psi(\phi(s),x,y,t)}\tilde r_1(s), \quad r_2(s) = e^{\psi(\phi(s),x,y,t)-\psi(s,x,y,t)} \tilde r_2(s)$$
which is a simple implication of \cite{N20b}.
The complex KP equation can reduce to the real KP-I or KP-II as follows.
Suppose that $u$ solves the complex KP equation.
If $u(x,y,t)$ is real for real values of $x,y,t$ then $u(x,y,t)$ solves the real KP-II equation.
However, if $u(x,iy,t)$ is real for real values of $x,y,t$ then $u(x,iy,t)$ solves the real KP-I equation. 

A solution to the complex KP equation produced in the manner might not reduce to a nonsingular real solution of either to the KP-I equation or the KP-II equation.
However, if we were to assume
$$\tilde r_1(s) \ge 0, \quad \tilde r_2(s) \le 0$$
are supported on some positive intervals and $\phi(\lambda)=-\lambda$, then these solutions reduce to primitive solutions to the KdV equation, which are real, smooth and bounded \cite{DNZZ20,N20a,NZZ19,DZZ16}.

For the choices
$$r(s) \ge 0, \quad r_1(s) \ge 0, \quad r_2(s) \le 0$$ supported on $$\mathcal{Q}=\mathcal{C} \subset \mathbb{R}^+,$$ and $\phi(\lambda) = -\lambda$, this construction gives a solution to the KdV equation. This reasoning is discussed in detail in \cite{ZM85,DZZ16,N20a,N20b}. For most other choice of embedding of $\mathcal{C}$ into $\mathbb{C}$, the construction will produce solutions to the KP equation. For $\mathcal{Q} = \mathcal{S}$, the construction will always produce a solution to the KP equation rather than the KdV equation since $\mathcal{S}$ can not be restricted to a one dimensional set.

Consider the case of $\mathcal{Q}=\mathcal{C}$ and a solution $u(x,t)$ to the KdV equation.
Let us consider the $t$ dependent family of one dimensional Schr\"{o}dinger operators
$$\hat H(t) = -\partial_x^2+u(x,t).$$
The (energy) spectrum $\sigma(\hat H) = \sigma(\hat H(t))$ is constant in $t$ and
$$\sigma(\hat H) = \{E=\lambda^2: \lambda \in \mathbb{R}, \; -i \lambda \in \mathcal{C}, \text{ or } i \lambda \in \mathcal{C}\}.$$
In other words, we can produce a potential with the above spectrum, and an explicit basis of (generalized) eigenfunctions of $\hat H$.
This is precisely the information we need to explicitly construct the spectral projection operators.

We can use intuition from finite and infinite gap theory and the trace formula to make some conjectures on the behavior of the solutions to the KP equation, the KdV equation and the inverse spectral theory of one dimensional Schr\"{o}dinger operators:

First, It seems likely that such a solution would either be quasi-periodic, or asymptotic to a quasi-periodic solution given the results in \cite{GGJM18,Sim}. It also seems likely that by associating gaps to the intervals $\mathcal{C}_n$, computing finite gap solutions, and then taking an infinite gap limit, would lead to an equivalent constructions of some solutions/potentials in the isospectral set of $u(x,0)$. 
Due to the fact that that gaps in the spectrum occurs at essentially all length scales smaller than the smallest interval $I$ such that $\mathcal{C} \subset I$, it is likely that the resulting solution has interesting multi-soliton interactions at all length scales smaller than $I$. These solutions to the KdV equation could therefore potentially exhibit complicated soliton gas dynamics.

Second, while the connection to spectral gaps is currently not clear for the distributions supported on Sierpinski gaskets, it still may be true that the Sierpinski surface discussed in this paper leads to solutions to the KP equation that have interesting multi-soliton interaction at all length scales. Further study into the Sierpinski surface and the nonlocal dbar problem could potentially lead to complicated soliton gasses when the support of the soliton spectrum is homeomorphic to $\mathcal{C}$. Even if soliton gasses of this type turn out to not be physically relevant, they may still be interesting as a completely solvable model of a soliton gas.

\end{document}